# Beyond Bilingual: Studying Quadrilingual Code-Switching with Conversational AI

Annika Bush[1,2], Melika Minnaar[3], Salys Sultan[4], MD Tasnimul Hasan Tashin[5], Ankur Bhatt[6], Meltem Aksoy[5]

[1]Human Geography, University of Bremen, Bremen, Germany
[2]Adrienne Arsht Community-Based Resilience Solutions Initiative, Smithsonian Tropical Research Institute, Panama City, Panama
[3]Avalon University, Willemstad, Netherlands Antilles
[4]SISSTEM Program, Faculty of Arts and Science, University of Aruba, Oranjestad, Aruba
[5]Research Center Trustworthy Data Science and Security, TU Dortmund, Dortmund, Germany
[6]TU Dortmund University, Dortmund, Germany

bush@uni-bremen.de, busha@si.edu, melika.ausom1597@avalonu.org, salys.sultan@ua.aw, tasnimulhasan88@gmail.com, ankur.bhatt@tu-dortmund.de, meltem.aksoy@tu-dortmund.de

**Abstract**

Multilingual speakers choose languages by topic and conversation partner. Large language models (LLMs) support these languages unequally. Research on code-switching with conversational AI has mostly examined bilingual pairs of well-resourced languages. In Curaçao and Aruba, many people use Dutch, English, Spanish, and the low-resource Creole Papiamentu/o. In a preregistered study, 42 university students held two five-minute chatbot conversations, one about university life and one about local food and culture. Surveys assessed language background, AI use, and switching comfort. 20 international students in Curaçao formed a comparison sample. Participants switched between turns and rarely mixed languages within a message. English dominated both conversations. Participants switched more often in the cultural conversation, where Papiamentu/o and Spanish rose from 15% to 24% of messages. Self-reports showed no lower comfort switching with the chatbot than with a person. Conversational agents must follow turn-by-turn language changes and handle local low-resource languages well.

## 1 Introduction

In every conversation, multilingual speakers decide anew which language to use. This depends on who they are speaking with, what they are talking about, and which language is expected of them in that particular situation [28, 47]. The same person can switch from one language to another within a matter of moments during a conversation covering various topics, such as work and personal life [49]. Similarly, multilingual speakers switch between languages within a single sentence; for example, when a word lacks a good equivalent in the language they are currently speaking [56].

One open question is how multilingual users deal with their multiple languages when interacting with Large Language Models (LLMs). Generative AI, particularly LLMs, has become a new type of conversation partner and faces unique challenges in meeting the needs of multilingual speakers. An LLM has only limited knowledge of different languages. The quality of its understanding and linguistic output depends on the pre-training corpus [42]. Error rates for languages with limited resources are much higher than for languages with extensive resources [36]. The same task posed in a language with fewer resources leads to measurably worse results [17]. Inputting in multiple languages is even more difficult [68]. The system therefore works to varying degrees of effectiveness depending on which language the user chooses for the interaction.

Most studies focus on switching between two languages, but few analyze user behavior when switching among three or more languages [60].

The Caribbean Small Island States Aruba and Curaçao offer one of the most suitable settings to study how multilingual speakers interact with LLMs. The residents of both countries (which are part of, or were formerly part of, the Kingdom of the Netherlands) are predominantly quadrilingual, meaning they use four languages in their daily lives: Dutch, English, Spanish, and their respective local Creole language. On Curaçao, this is Papiamentu [18], on Aruba it is Papiamento [37]. These five languages differ sharply in how well language technology supports them. English and Spanish are among the best-resourced languages, and Dutch follows one class below [34]. Papiamentu and Papiamento are low-resource languages that major translation systems added only in 2022 (Papiamento) [50] and 2024 (Papiamentu) [25]. Two circumstances limit the resources of both languages: (1) The speaker base is very small, with estimates ranging from about 250,000 to 350,000 speakers worldwide [21, 32], and (2) although both languages sound very much alike, their written orthographies differ significantly from one another, which makes the training corpus per language even smaller.

Computer-assisted research on code-switching focuses primarily on bilingual language pairs [13, 20, 51, 66]. Furthermore, the language pairs mostly come from the upper end of the resource gradient [16, 19] and are based on self-reported data from the participants. We are not aware of any study that has observed multilingual users using four languages in conversation with an LLM, one of which is a Creole language—a category that most systems have only recently begun to cover.

This study therefore asks: *How do multilingual speakers use their multiple languages when interacting with an LLM?* We divided this main research question (RQ) in four sub-questions:

**RQ1** Which languages do participants use with the chatbot, and how often do they move between them?

**RQ2** Does the same participant use different languages when the topic shifts from an institutional to a cultural domain?

**RQ3** At which points in the exchange do switches occur?

**RQ4** How comfortable are participants switching languages with a chatbot?

To answer these questions, we conducted an interaction study with $N = 42$ students from the University of Curaçao and the University of Aruba, along with a comparison sample of $N = 20$ international students studying in Curaçao. The participants completed two five-minute chatbot conversations about two different topics as well as two surveys.

## 2 Literature Review

### 2.1 Code-Switching as a Socially Situated Practice

Gumperz [27, p. 1] defines code switching as ‘the juxtaposition of passages of speech belonging to two different grammatical systems or subsystems, within the same exchange’. The shift between languages follows certain patterns. Gumperz [28] describes this shift as the fulfillment of recognizable interactive functions, including the specification of the addressee, repetition, emphasis, and the marking of attitude. He distinguishes between situational and metaphorical code switching. In situational code-switching, the instances of code-switching are triggered by a change in situation. In metaphorical code switching, the speaker uses a different language to introduce meanings or associations associated with that language into the existing situation. According to this view, code-switching is a communicative resource, and its placement conveys information about how a speaker interprets the moment.

Myers-Scotton [47] further conceptualizes code choice as part of a system of social meaning. Within a given interaction, speakers typically have an unmarked code choice that corresponds to the expected social relations and situational norms. A switch to another code can constitute a marked choice and thereby signal a shift in interpersonal relations, social distance, or communicative stance. Which language code is considered unmarked depends more on the context than on the speaker, and in multilingual communities, the dominant or institutional language may occupy this position even if it is not the native language of those present [1]. Bhatt and Bolonyai [7] reframe these observations as a small set of sociocognitive constraints on multilingual competence, from which the local functions can be derived.

Two factors are crucial for the use of unmarked codes. The first is the conversation partner. Speakers either adapt to their partner's language use or deviate from it in order to manage alignment and social distance [24, 57]. The nature of the interaction context determines how willingly and at what cost a speaker switches codes within that context [39]. The second factor is the topic. When modeling language switching in Nigerian online discussions, Ndubuisi-Obi et al. [49] notes that topic-related variations operate in parallel with audience-related factors. Where languages are associated with different spheres of life within a community, the topic under discussion is part of the context that determines which language code is unmarked.

A generative AI chatbot is a new type of conversation partner, and the two scenarios above support opposite expectations about it. If no unmarked code has yet been established for such a conversation partner, the choice of language should be responsive and track the topic and the agent's own language. If, however, an unmarked code has already been transferred from the surrounding technological environment, where interfaces commonly default to a single dominant language [14], one language may be expected to prevail irrespective of the topic.

## 2.2 Switching in Written and Digital Interaction

Most theories of code-switching are based on spoken language, but specific patterns can also be identified in written language. Themistocleous [62] shows that the shift between Cypriot and Standard Greek in online chats is shaped both by macrosociolinguistic relationships between the lingual varieties and by the possibilities offered by the digital medium, and that users exploit this to present different identities within a single conversation. Androutsopoulos [3] considers computer-mediated code-switching, for these reasons, to be a distinct field of research. Multilingual online practice is not tied to competence in the traditional sense, since authors draw on their entire repertoire rather than on languages they supposedly speak [8], and much of what happens in digital contexts could be described as translingual code play [9, 22].

The functions described for the shift in written language use largely correspond to those documented in spoken language. Begum et al. [6] developed an annotation scheme to analyze language-switching between Hindi and English on Twitter. They drew on established categories from the literature on spoken language, which they applied to the written data, and found that the majority of these categories are transferable. Studies on other languages have shown that language switching is used for emphasis, humor, solidarity, to fill lexical gaps, and to select an audience [23, 38, 63]. Written interaction adds its own resources, as spelling choices, punctuation, and emojis convey part of the meaning and do not serve merely decorative purposes [38, 48].

Nearly all of the written data analyzed from earlier studies comes from public or semi-public online platforms where an audience is present and where language choice serves an identity-forming function in the presence of that audience. Androutsopoulos [4] shows that when different target audiences merge into a single one, authors adjust their language choice accordingly.

## 2.3 Multilingualism in Aruba and Curaçao

Papiamentu emerged from Curaçao in the 17th century and has been the main language spoken ever since across all social classes and ethnicities [33]. From there, it spread to Aruba and Bonaire. Both Creoles are mainly the same spoken language, and language identification resources assign both the same ISO 639-3 code [61]. However, the islands write the languages differently. Aruba's Papiamento is based on etymological orthography, while Curaçao's Papiamentu has a phonological base, a split that followed the political separation of Aruba from the Netherlands Antilles [21].

Due to the island's colonial history, Dutch remains an official language, but just under 10% of the local population speaks it on a daily basis [33]. In addition, Spanish is present due to immigration and tourism, and English rose to a centrally important fourth language in the 20th century, especially in Aruba, because of tourism [30]. People living in Aruba and Curaçao usually know and use all four languages, depending on the environment. While Dutch is still the language of administration, people tend to prefer Papiamento and other languages at home as seen in Table 1 for Aruba. Also in Curaçao, Papiamentu dominates as main language at home with 78.8%, followed by Spanish (8.4%), Dutch (7.9%) and English (3.8%) [11].

**Table 1.** Census 2020: Main Languages Spoken at Home in Aruba [10].

| Language | National Average (%) |
|---|---|
| Papiamento only | 43.2 |
| Papiamento & Spanish | 18.9 |
| Papiamento & English | 9.4 |
| Papiamento & Dutch | 6.8 |

The local education systems show linguistic differences between the islands. In Curaçao, Papiamentu is the language of instruction in elementary school, while in Aruba, teachers use Dutch; there, Papiamento is taught only as a subject [18]. From secondary school onward, students are taught in Dutch on both islands, and at the university level, English becomes another dominant language of instruction. There are ongoing discussions about which language should dominate in education [40] and which languages are suitable for academic achievement [46]. This process of negotiation also takes place within the classroom, where students and teachers use their different languages to affirm and question what is considered appropriate [65]. Papiamentu and Papiamento also serve as identity-forming elements: They are the most commonly used languages in private life, and attitudes toward it remain positive, despite the increasing prevalence of English and Spanish [37].

## 2.4 Code-Switching and Language Technology

LLMs do not serve all languages equally. Li et al. [42] report a strong correlation between per-language performance and a language's share of the pretraining corpus, and find that low-resource languages occupy a narrow region of the embedding space relative to high-resource ones. Kazemi et al. [36] found that the availability of digital resources in a language accounts for 44% of the variance in GPT-4o's alignment with survey-measured societal values and 72% for GPT-4-turbo, and error rates for low-resource languages run more than five times those for high-resource ones [36]. The gap is visible in ordinary use as well, since prompts written in English outperform prompts written in Bangla, Hindi and Urdu on the same tasks across GPT-4, Llama 2 and Gemini [17].

The relationship is regular enough to be modelled. Scaling laws predict multilingual performance from how pretraining data is allocated across languages [29], and adaptive formulations extend this to transfer between specific pairs [43]. Because allocation differs between models, two systems represent two different allocation regimes. Performance is nonetheless not a simple function of volume. Adding multilingual data improves low-resource language modelling by an amount comparable to enlarging the monolingual corpus by up to a third, with larger gains when the added languages are syntactically similar, while degrading performance for high-resource languages [12]. An imbalanced training distribution can outperform a balanced one [58]. And much of the observed disparity shrinks once segmentation and data exposure are held constant, which locates it in tokenisation, sampling and evaluation design rather than in the languages themselves [59].

Mixed-language input remains harder than either language alone. Multilingual models still handle code-switched text poorly [68], and the resources used to evaluate them are narrow in a specific way. Across 68 code-switching datasets, most pair a language with English, and collection practices largely ignore variation by location, speaker background and register [19]. A recent survey of code-switching research in the LLM era similarly finds a strong concentration on high-resource language pairs, while low-resource and non-English combinations remain underrepresented [60]. Model capability is thus understood as a gradient, while code-switching is studied in pairs, usually with English at one end. This gap is particularly pronounced beyond bilingual settings. Research on switching across three or more languages remains sparse, leaving multilingual interaction patterns that extend beyond conventional language pairs poorly understood [60].

Joshi et al. [34] sort languages into six classes by counting the resources available for each: Wikipedia articles on the unlabeled side, Linguistic Data Consortium (LDC) and European Language Resources Association (ELRA) catalog entries on the labeled side. English and Spanish fall in the top class (nr. 5, 'The Winners'), Dutch in the 4th class ('The Underdogs'). Papiamentu is categorized into class 1, 'The Scraping-Bys' (there is no mention of Papiamento as a distinct language). NLLB-200 ("No Language Left Behind") draws its cutoff at one million publicly available sentence pairs aligned with another language; below this threshold, a language is not considered resource-rich [15, 50]. Papiamento fell below this threshold and was classified as resource-poor when the project added it [50] (Papiamentu is not listed), whereas English, Spanish, and Dutch are well above the threshold under both counting methods. Therefore, it is to be expected that LLMs handle Papiamentu/o much worse than the other three languages, and English and Spanish better than Dutch.

## 2.5 Multilingual Users and Conversational Systems

Most studies focus on how systems perform on multilingual data but only a small body of work examines how multilingual people behave with interactive systems. Bawa et al. [5] find that multilingual users prefer chatbots capable of code-mixing and that nudging users toward mixed input is a low-risk design policy. Ahn et al. [2] compare code-switching strategies for dialogue agents and find that users respond poorly to mixing that is unpredictable or ungrammatical. Parekh et al. [54] found that users accommodate to an agent's code-switching style and rate code-mixing agents as more natural than monolingual ones. Cihan et al. [14] document that voice assistants are monolingual by default and that this constrains the bilingual user experience. Choi et al. [13] similarly found that multilingual users perceived conversational agents as monolingual conversation partners and adjusted their own language use to what they expected the agent to understand. Participants avoided using languages they did not expect the agent to handle and adapted their speech following unsuccessful interactions with the system. Outside dialogue, Wang and Komlodi [64] show that multilingual users searching for information switch languages deliberately, to obtain translations and to reach domain-specific material. Surveys of the field note that the social determinants of switching are precisely what language technologies fail to represent [20]. Almost all of the existing research concerns two languages, usually with English as one of them; where a third or fourth language appears, it remains peripheral to a dominant pair [16].

It remains unclear what this research means for user comfort, and the two lines of evidence point in opposite directions. The preference for systems that mix different languages [5] suggests that users welcome mixed interaction and would switch between languages at least as freely as they would with a human conversation partner. In contrast, assistants operate in a single language by default [14], and error rates rise sharply for languages with fewer available resources [36], so a user who expects a system to master only a portion of their linguistic repertoire has reason to limit themselves to the portion that the system does master.

The languages involved in previous studies cluster near the top of the resource gradient, so they do not cover what multilingual speakers do when part of their repertoire is barely represented. Additionally, the evidence comes either from manipulating what the system does, or from self-report and retrospective logs. So far, research mostly focuses on existing data corpora or self-reported data and is lacking induced interactions [49, 55].

We investigate how students use their multiple languages when talking to a system that inherently handles them very differently.

## 2.6 Hypotheses

The literature review leads to four hypotheses.

**H1:** *Multilingual students will switch languages frequently when interacting with an AI chatbot.* Switching is a patterned communicative resource [28], and the repertoire available to speakers on these islands is both large and in daily use across separate domains [18, 30, 33]. The typed medium does not suppress switching: alternation is well documented in online chat and other written interaction [3, 62], and writers there draw on their full repertoire and not only on languages they would claim to command [8]. Multilingual users also report a preference for agents that can code-mix [5]. Taken together, these give us reason to expect frequent alternation in this setting.

**H2:** *Participants will use predominantly Dutch and English in the university-related conversation.* Bilinguals treat one code as the unmarked choice for a given interaction type [47]. The topic is part of the setting that fixes the unmarked choice [28, 49]. In Curaçao and Aruba, Dutch and English are the principal languages of university instruction [18, 30, 40]. A conversation about studies and career options should therefore invoke the institutional domain and the codes indexed to it.

**H3:** *Participants will use more Papiamentu/o and Spanish in the culture-related conversation.* The same reasoning applied to the opposite domain. Papiamentu is the language of all social classes on the island [33] and the primary marker of local identity [37, 65], while Spanish is present through immigration and tourism [30].

**H4:** *Participants will report greater comfort switching languages with a human than with an AI chatbot.* Model quality is dependent on the share of the language's pretraining corpus [42]. Papiamentu and Papiamento are both low-resource languages, and mixed-language input is harder still [67]. Assistants operate monolingually by default [14], and users respond poorly to agent mixing that is unpredictable or ungrammatical [2]. A speaker who expects a system to handle only part of their repertoire has reason to restrict themselves to the part it handles. We assume that users who expect a system to struggle will also feel less comfortable using their full repertoire with it. Previous evidence asked what users want the system to do [5], H4 asks what users themselves are comfortable doing.

**Scope.** H1 to H3 concern behavior observed in the transcripts; H4 concerns self-report. Analyses beyond these four hypotheses, including the trigger taxonomy, analyses of the comparison sample, and all model-stratified contrasts, were not preregistered and are reported as exploratory.

## 3 Method

This study has been preregistered before data collection: https://aspredicted.org/4s3ux6.pdf.

### 3.1 Study Design

Curaçao and Aruba provide a unique natural setting for studying multilingual human–AI interaction. Unlike controlled laboratory studies that investigate predefined bilingual language pairs, multilingual communication is part of everyday life on both islands. University students routinely move between English, Dutch, Spanish, and Papiamentu/o depending on the conversational context, communication partner, and social setting. Because these languages occupy markedly different positions on the language-resource spectrum of contemporary LLMs, this setting allows us to observe multilingual AI interactions without constraining participants' language choices.

The study was conducted through a custom web-based application that integrated the participant information sheet, informed consent form, multilingual questionnaires, and chatbot interface into a single workflow. The application linked survey responses and chatbot conversations through anonymized participant identifiers.

Following informed consent, participants completed the study protocol in a fixed sequence: a university-related chatbot conversation (Chat1), a survey (Survey 1), a culture-related chatbot conversation (Chat2), and another survey including demographic data (Survey 2). This order was identical for all participants. The complete study took approximately 20 minutes, including two five-minute chatbot conversations.

To accommodate the multilingual study population, the informed consent form, and survey questionnaires were available in English, Dutch, Papiamentu/o, and Spanish. Participants could complete these materials in their preferred language, independently of the languages they used during the chatbot conversations. An overview of the study procedure is shown in Figure 1.

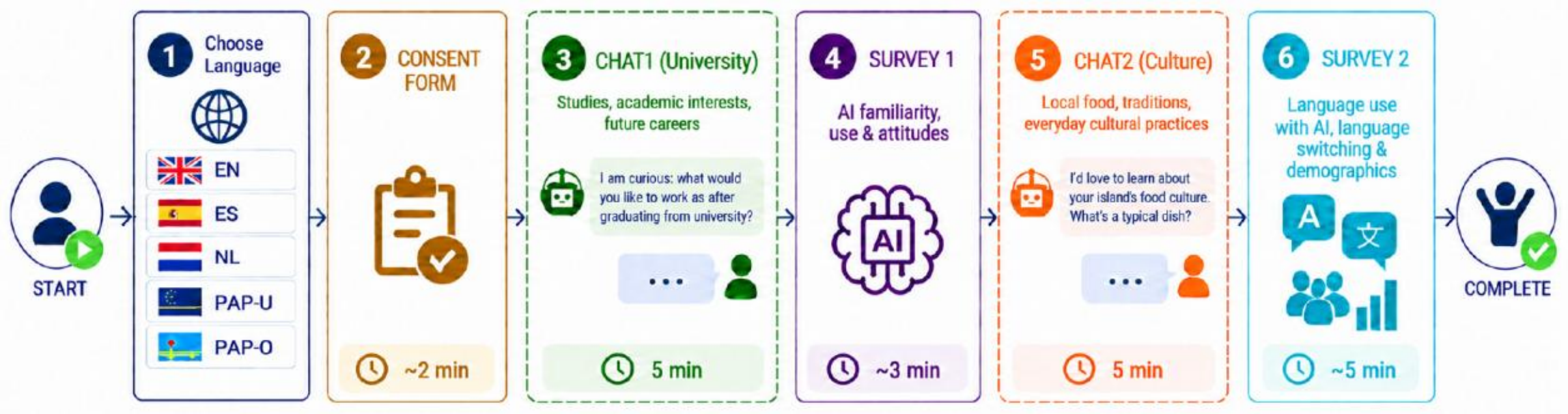

**Figure 1.** Overview of the study procedure. Participants selected their preferred study language, provided informed consent, and completed two surveys interleaved with two chatbot conversations. Chat1 focused on university-related topics, whereas Chat2 focused on local culture and everyday practices. PAP-U denotes Papiamentu (Curaçao), and PAP-O denotes Papiamento (Aruba).

### *3.1.1 Survey*

Here, we only describe the measures used in the analyses below. Participants completed two surveys (Appendix A): The consent form and all surveys were available in English, Dutch, Papiamentu/o, and Spanish.

Survey 1 was filled out in between both chatbot interactions. It assessed *AI familiarity* with a single five-point Likert scale (*not familiar at all* (1), *very familiar* (5)) and frequency of purposeful AI use (*never* (1) to *daily* (5)). It also included the four-item Artificial Intelligence Attitude Scale (AIAS-4 [26]) on a six-point agreement scale.

Survey 2 was placed after the second chatbot interaction. It asked participants to rate their proficiency in each language of their repertoire on a 5-point Likert scale. Following the preregistration, a *multilingualism index* was calculated by counting the number of languages rated 3 or higher. Participants also named the languages they mainly speak with family, with friends, and at university, and ranked the languages they use with AI systems.

*Relative switching comfort* asks whether participants feel more comfortable switching languages with an AI chatbot than with a person, answered on a five-point Likert scale in two items (e.g., "I feel more comfortable switching languages when interacting with a person than with an AI chatbot.")

*Self-reported switching behavior* was measured with seven items, asking how often participants switch languages within a single chatbot conversation (e.g. "I switch to another language if I cannot recall a word or expression."), on a five-point Likert scale from *never* (1) to *always* (5), with an additional *not applicable* option. Item 1 ("I stay in one language throughout the conversation") was reverse-scored before aggregation. The composite is the mean of the seven items, with higher values indicating more frequent self-reported switching.

Survey 2 closed with demographic questions on country of origin, age, gender, and field of study.

### *3.1.2 Chatbot interaction*

The two chatbot conversations were designed to represent communicative contexts associated with different language practices on the islands and to examine whether conversation topic was associated with participants' language choices. Chat1 focused on university-related topics, such as participants' studies, academic interests, and future careers. Because higher education in Curaçao and Aruba is primarily conducted in English and Dutch, this context was expected to encourage the use of these institutional languages. Chat2 focused on local food, traditions, and everyday cultural practices, where greater use of Papiamentu/o and Spanish was expected given their role in local identity and everyday communication.

Both conversations began with the same English opening message to provide a consistent starting point. Participants were informed that the chatbot did not know their preferred language and that they were free to respond in any language and switch languages as they would during everyday chatbot use. No further instructions regarding language choice or language switching were provided, allowing multilingual behaviour to emerge naturally. After each participant's first response, the chatbot was instructed to adapt to the participant's language use without explicitly commenting on their language choice.

**Chatbot Models.** Participants interacted with an AI chatbot powered by one of two LLMs: GPT-5.5 [53] and Llama 4 Maverick [45]. We configured both models with identical system prompts, conversation scenarios, and generation parameters (temperature = 0.7, maximum output length = 300 tokens). To encourage natural dialogue while maintaining comparable interactions across participants, chatbot responses were limited to 1–3 sentences.

Participants were not randomly assigned to the chatbot models. The model available during each data-collection session determined whether participants interacted with GPT or Llama, resulting in an opportunistic between-subjects design. Consequently, chatbot model and participant-group composition are confounded, and the study was not designed to estimate causal effects of model choice. The model-stratified analyses therefore describe patterns within each chatbot group rather than test whether GPT and Llama produced different participant behaviours.

Because current LLMs have been shown to infer the expected output language unreliably when it is left implicit in code-switched input [51], we did not rely on the chatbot to make this inference on its own. Instead, the system prompt explicitly instructed the model to detect the language of the participant's most recent message and reply entirely in that language. This design choice was intended to ensure that the language-switching patterns we observe reflect participants' own language choices rather than inconsistencies in the chatbot's language selection. The system prompts used for both conversation contexts are provided in Appendix C.

We selected GPT-5.5 and Llama 4 Maverick to represent two distinct model ecosystems: GPT-5.5 as a proprietary, closed-weight model and Llama 4 Maverick as an open-weight model. Both model families support multilingual interaction, making them suitable for examining language use and language switching in a multilingual setting. We refer to GPT-5.5 as *GPT* and Llama 4 Maverick as *Llama* throughout the remainder of the paper.

## 3.2 Pilot Study

Before the main data collection, we conducted multiple pilot studies to evaluate both the technical functionality of the study platform and the suitability of the study materials. Initial pilot sessions were performed by members of the research team to identify technical issues related to the multilingual questionnaires, chatbot interface, and data logging. Subsequently, local multilingual speakers from Curaçao completed the study and provided qualitative feedback regarding the clarity of the questionnaires, the naturalness of the conversation scenarios, and the quality of the multilingual translations. Based on this feedback, several refinements were made to the web application, chatbot prompts, survey wording, and user interface before the main data collection commenced.

## 3.3 Data Collection

Data collection was conducted in two phases. The primary sample consisted of university students recruited from the University of Curaçao and the University of Aruba between April and June 2026. These institutions represent the multilingual educational environment characteristic of Curaçao and Aruba, where students routinely use English, Dutch, Papiamentu/o, and Spanish in both academic and everyday communication. Data collection was carried out in collaboration with local academic partners at the participating institutions, who facilitated participant recruitment in university seminars. *[Additional information about the researchers was blinded for review.]*

For comparison, we subsequently collected a second dataset from Avalon University School of Medicine in Curaçao using the identical study protocol. Unlike the primary sample, which consisted of mostly local students embedded in the multilingual educational environments of Curaçao and Aruba, Avalon primarily enrolls international medical students and conducts all teaching in English. No local students from Curaçao,

Aruba, or Bonaire participated in the comparison sample. The students came from different linguistic backgrounds and were free to use any language during the chatbot interactions.

The comparison sample therefore allowed us to examine whether the language-switching patterns observed among students embedded in the multilingual educational environments of Curaçao and Aruba also emerged among students living in Curaçao but studying in a predominantly English-medium international university environment.

### 3.4 Ethics

The study was approved by the Ethics Committee. Before participation, all participants received a multilingual participant information sheet describing the study objectives, procedures, data handling practices, and their rights as participants. Written informed consent was obtained electronically before data collection. Participation was voluntary, and participants could withdraw at any time without providing a reason. All survey responses and chatbot conversation logs were anonymized before analysis, and no personally identifiable information was retained in the analytical dataset.

### 3.5 Data Preprocessing

Raw survey responses and chatbot conversation logs were collected across multiple data collection sessions. Before analysis, data from these sessions were processed and combined to construct unified survey and conversational datasets.

For the survey data, the two surveys completed during each session were first merged using anonymized participant identifiers. The resulting datasets from different collection sessions were then combined into a single survey dataset. Variable names were harmonized across survey versions, and duplicate and test records were removed. Only participants who completed two surveys and both chatbot conversations were retained in the final sample. Multilingual survey responses were preserved in their original languages during preprocessing and converted to numerical scores where required for subsequent analyses.

For the conversational data, chatbot logs collected across different sessions were merged while preserving participant identifiers, conversation identifiers, timestamps, message order, and speaker roles. Each conversation was annotated with the corresponding chatbot model and conversation topic. Research accounts, pilot sessions, and test participants were removed before analysis. Messages were then chronologically ordered within each participant and conversation to reconstruct the original dialogue sequence.

The same preprocessing procedure was applied separately to the primary and comparison datasets.

### 3.6 Language Identification

Reliable language identification represents a methodological challenge in multilingual conversations involving low-resource languages such as Papiamentu/o [66]. Widely used language identification tools, such as *langid.py* [44], operate on predefined sets of supported languages that do not include Papiamentu/o. More recent broad-coverage tools, such as GlotLID [35], extend support to more than 1,600 languages, including low-resource languages, using a FastText-based classifier and include an ISO 639-3 label for Papiamento. However, standard text-level language identification with such tools does not distinguish between the Papiamentu (Curaçao/Bonaire) and Papiamento (Aruba) orthographic variants and does not directly identify the presence of multiple languages within a single message, as required for our within-message mixing analysis. Consequently, relying on these tools could lead to inaccurate identification of these languages and systematically underestimate multilingual behavior in our study.

We therefore used an LLM-as-a-judge approach for language annotation, employing GPT-5.4-mini [52] to classify the language content of each message. Every user and chatbot message was submitted independently using a deterministic prompt (temperature = 0). For each message, the model identified the dominant language and indicated whether multiple languages were present. Within-message language mixing was defined as the presence of more than one language within a single participant message; an illustrative example is provided in Table 2. When within-message mixing was detected, the model returned all languages present in the message. The resulting annotations were manually reviewed by the authors, who corrected misclassified language labels and within-message mixing classifications where necessary and resolved uncertain cases through discussion. These annotations were then used to identify both dominant-language changes across messages and within-message language mixing. The annotation prompt is provided in Appendix D.

During annotation, we initially distinguished between Papiamentu and Papiamento based on their regional orthographic conventions to identify possible differences exploratory. However, many messages were too short or contained insufficient orthographic cues to reliably distinguish between the two variants. Since we preregistered to focus on the local Creole in relation to the other three languages (English, Dutch, Spanish) anyway (without differentiating between Papiamentu and Papiamento), we merged both labels into a single Papiamentu/o (`pap`) category for all subsequent analyses.

**Table 2.** Illustrative example of within-message language mixing.

| Languages | Participant message |
|---|---|
| PAP + NL | Mi lo ke trahá komo dosente di eseñansa di fundeshi i kisas despues di algun aña komo kabes di skol òf den *beleid/onderwijs inspectie*.<br>*("I would like to work as a foundation-education teacher and maybe, after a few years, as a school head or in 'policy'/education inspection.")* |

*Note.* PAP = Papiamentu/o; NL = Dutch.

## 3.7 Language Switching

Language-switch analyses focused exclusively on participant messages because the objective of this study was to investigate users' language choices rather than chatbot behaviour. A language switch was defined as a change in the dominant language between two consecutive participant messages within the same conversation. Messages were first ordered chronologically, after which the dominant language of each message was compared with that of the immediately preceding participant message. Whenever the dominant language changed, the message was marked as a language switch. For each detected switch, we additionally recorded the previous language, new language, conversation topic, switch order, and switch direction within the conversation. This procedure enabled analyses of both overall switching frequency and specific transition patterns (e.g., English → Papiamentu/o).

## 3.8 Trigger Classification

Although switch frequencies quantify multilingual behavior, they do not explain why participants change languages. To identify conversational triggers associated with language switching, we developed a trigger taxonomy tailored to multilingual human–AI interaction. We used GPT-5.4-mini [52] as an LLM-as-a-judge to classify the conversational trigger associated with each detected language switch. For every switch, the model received three consecutive messages: (i) the participant message immediately preceding the switch, (ii) the chatbot response, and (iii) the participant message following the switch. Based on this conversational context, the model assigned one of five predefined trigger categories. The resulting classifications were independently reviewed by two human reviewers, with inter-rater agreement on the trigger labels assessed using Cohen's κ (κ = 0.82). The five trigger categories, their definitions, and illustrative examples from the corpus are presented in Table 3. The classification prompt is provided in Appendix E.

**Table 3.** Trigger taxonomy for language switches with illustrative examples from the corpus. Language tags: EN English, NL Dutch, ES Spanish, PAP Papiamentu. The arrow marks the participant message on which the switch was recorded. English translations are given in italics.

| Nr. | Trigger | Definition | Example |
|---|---|---|---|
| 1. | AI New Question | The chatbot introduced a new question that was followed by a language change. | P159 [EN]: I usually prefer mild, flaky fish like Red snapper or Mahi-mahi<br>AI [EN]: Do you season the fish simply, or with a special family-style marinade?<br>→ P159 [NL]: Met een speciale familiemarinade. *("With a special family marinade.")* |
| 2. | AI Language Mismatch | The chatbot responded in a different language from the participant's previous message. | P95 [EN]: I eat it here and there<br>AI [NL]: Maar heb je ook een speciale herinnering aan het eten van bonchi? *("But do you have a special memory of eating bonchi?")*<br>→ P95 [NL]: Ik eet zo vaak bonchi dat ik er geen speciale herinnering aan kan koppelen, nee *("I eat bonchi so often that I can't attach a special memory to it, no.")* |
| 3. | User Explicit Request | The participant explicitly requested changing the conversation language. | P152 [PAP]: Mi gusta hot dog i pastechi di keshi mas tantu *("I like hot dogs and cheese pastechi the most.")*<br>AI [ES]: ¿Lo preparas tú mismo o prefieres comprarlo en algún lugar especial? *("Do you make it yourself or prefer to buy it somewhere special?")*<br>→ P152 [EN]: Speak Papiamentu! |
| 4. | AI Misunderstanding | The chatbot misinterpreted the participant's message or shifted the conversation away from the participant's intended meaning, leading a language switch. | P87 [NL]: Huidige situatie *("Current situation.")*<br>AI [MIXED]: E huidige situatie, ta un tema interesante, kiko ta bo ta mira pa cambia den e mundo di ayo, ora bo bira un maestro? *("The current situation is an interesting topic. What would you like to change in today's world when you become a teacher?")*<br>→ P87 [EN]: I didn't understand the question. |
| 5. | Unexplained | Code-switching with no detectable external trigger. The participant simply chose a different language without an identifiable reason in the surrounding conversational context. | N/A (we could not find an unexplained language switch in our dataset but included the category in our analysis to show we are aware that there could be other cases than the above-mentioned) |

## 3.9 Analytic Approach

To evaluate H1–H3, we examined participants' language-switching behavior and language use across the two conversational contexts. Participants interacted with either Llama ($n$ = 24) or GPT ($n$ = 18), and no participant used both systems. Analyses of H1–H3 were conducted across the full sample as well as separately within these two chatbot groups. H4 was evaluated across the full primary sample using participants' self-reported comfort measures.

Normality of the relevant paired differences was assessed using Shapiro–Wilk tests to select between paired $t$-tests and Wilcoxon signed-rank tests on a comparison-by-comparison basis. Where multiple related comparisons were conducted, Holm's step-down correction was applied to control the family-wise error rate at $\alpha = .05$ [31]. We report Holm-adjusted $p$-values for these comparisons. Because different statistical tests were used where appropriate, effect sizes are reported as Cohen's $d$ for paired $t$-tests and rank-biserial $r$ for Wilcoxon signed-rank tests. These effect-size measures are interpreted within each analysis and are not compared directly across chatbot groups.

# 4 Results

## 4.1 Sample Characteristics

After preprocessing, the primary sample consisted of $N = 42$ participants who completed two surveys and both chatbot conversations. As shown in Figure 2, the participants represented a diverse multilingual student population with respect to gender, country of origin, field of study, and age. Most participants were from Curaçao (69.0%) or Aruba (19.0%), reflecting the primary recruitment settings of the study. Their multilingual language repertoires and language use across different communication contexts are summarized in Figure 3. Language preferences varied markedly by context, with Papiamentu/o most commonly reported for family and friendship contexts, Dutch for university settings, and English for interactions with AI.

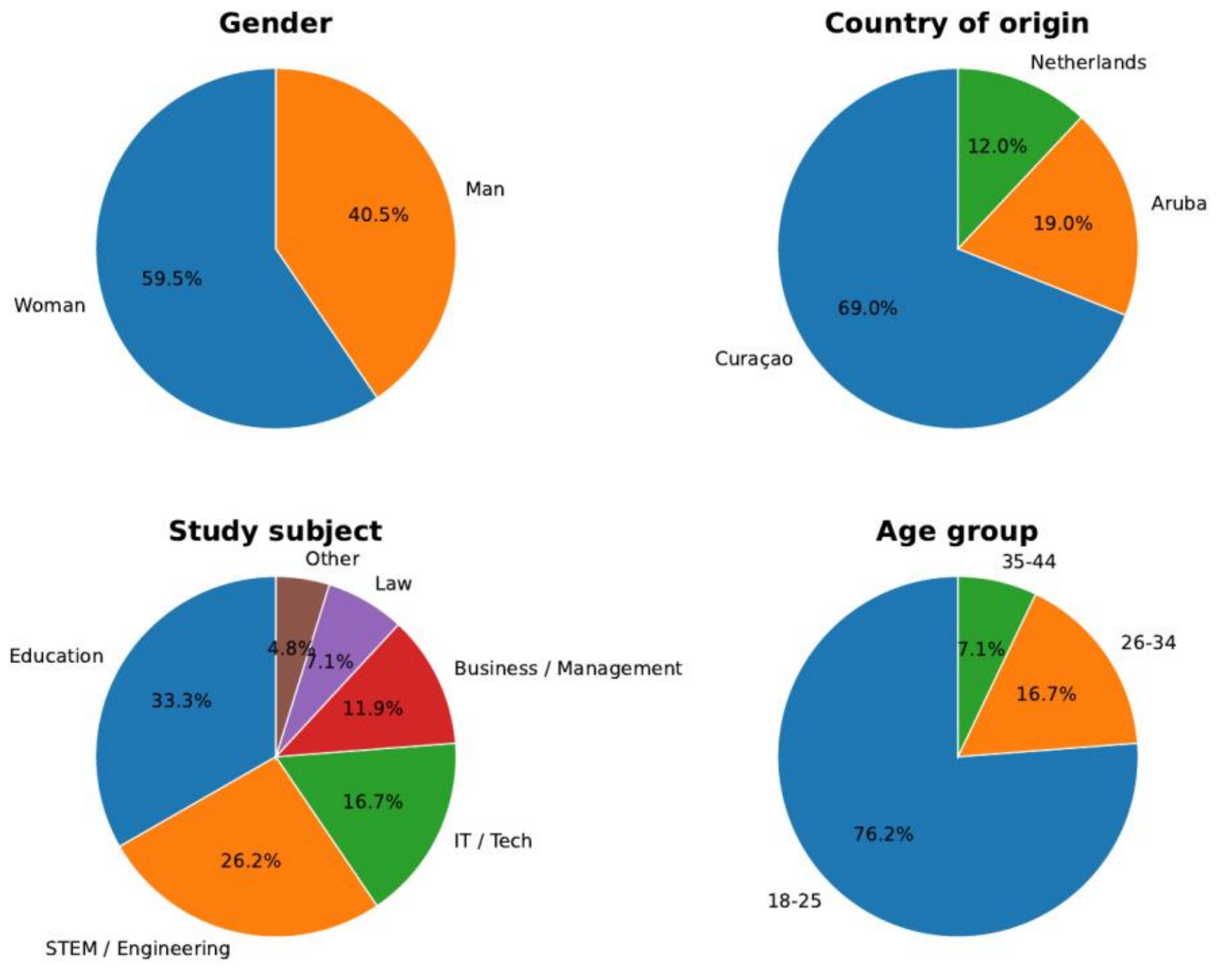

**Figure 2.** Participant demographics for the primary study sample ($N = 42$), showing the distributions of gender, country of origin, study subject, and age group.

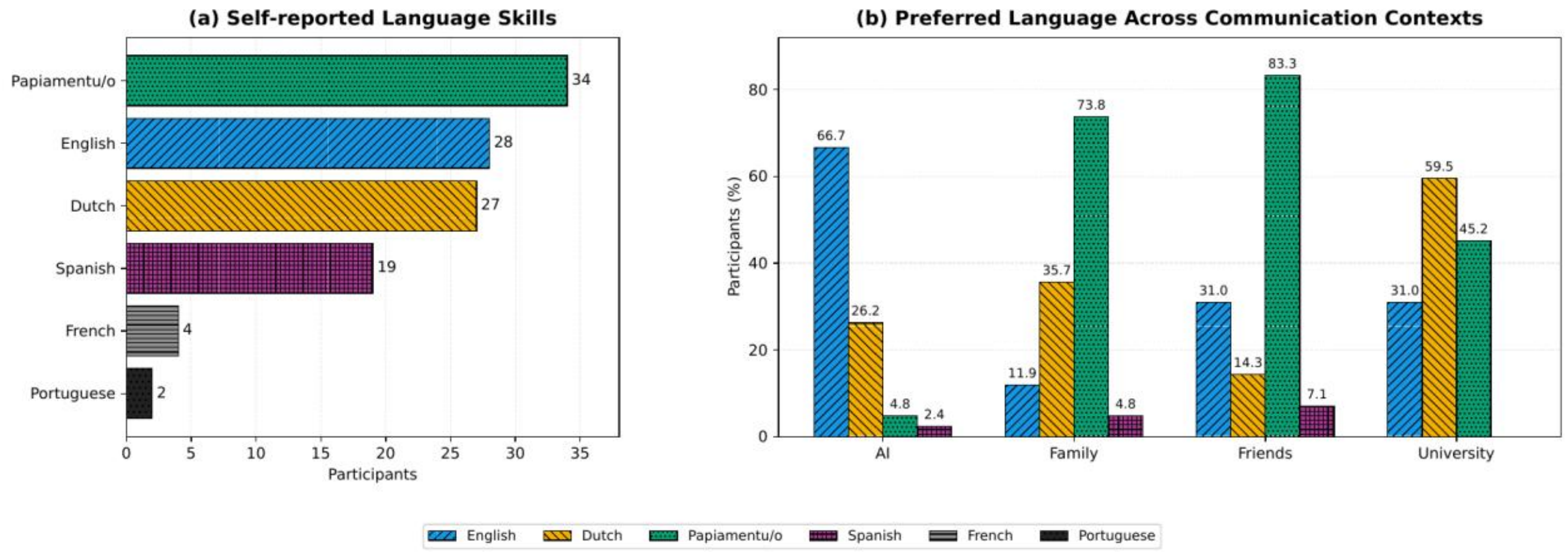

**Figure 3.** Language background of the primary study sample ($N = 42$). (a) Self-reported language skills across all participants. (b) Preferred language across different communication contexts.

Participants reported moderate to high familiarity with AI systems ($M = 3.64$, $SD = 0.96$), with no difference between the GPT ($M = 3.61$, $SD = 0.92$) and Llama ($M = 3.67$, $SD = 1.01$) conditions ($U = 214.5$, $p = .979$, $r = -.01$).

Most participants used AI systems daily (59.5%) or weekly (31.0%); three used them monthly and one never. Use frequency did not differ significantly between the GPT ($M = 4.22$, $SD = 1.00$) and Llama ($M = 4.63$, $SD = 0.65$) conditions ($U = 161$, $p = .112$, $r = -.25$).

## 4.2 Interaction Characteristics

Table 4 summarizes the interaction characteristics across the two conversational contexts and chatbot groups. User messages were consistently shorter than chatbot responses. The mean number of user turns was similar across the two conversations in the Llama group (5.2 in both), whereas the GPT group showed more user turns in Chat2 than in Chat1 (7.3 vs. 5.4).

**Table 4.** Descriptive characteristics of the chatbot interactions.

| Dataset | *N* | User msgs. | Mean turns | User words/msg. | AI words/msg. |
|---|---|---|---|---|---|
| Chat1 Llama | 24 | 125 | 5.2 | 7.8 | 21.8 |
| Chat2 Llama | 24 | 126 | 5.2 | 5.8 | 21.9 |
| Chat1 GPT | 18 | 97 | 5.4 | 11.4 | 25.3 |
| Chat2 GPT | 18 | 131 | 7.3 | 8.4 | 23.7 |

## 4.3 Hypothesis Testing

### *4.3.1 Language Switching Frequency (H1)*

As shown in Figure 4, language switching occurred across both conversational contexts, but its frequency varied considerably across participants. In the Llama group 54% switched at least once in Chat1 and 42% in Chat2. In the GPT group, 28% switched at least once in Chat1 and 67% in Chat2. Median switch counts ranged from 0 to 1.5. Switching was therefore clearly present, but frequent switching did not characterize participants consistently across conversational contexts. H1 received partial descriptive support.

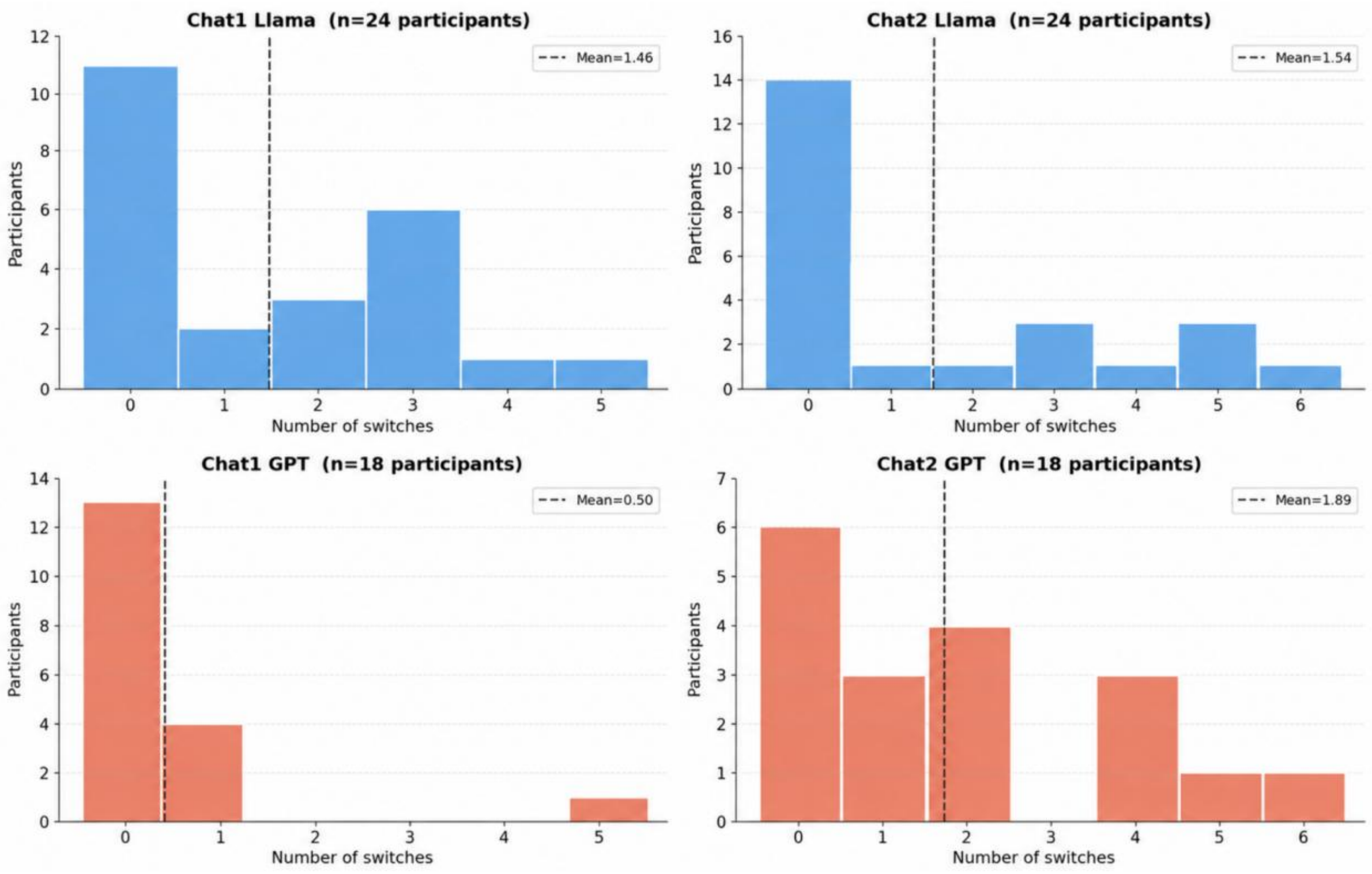

**Figure 4.** Distribution of the number of language switches per participant across the university-related (Chat1) and culture-related (Chat2) conversations for the Llama and GPT groups.

### *4.3.2 Language Choice by Topic (H2, H3)*

Figure 5 shows a contextual pattern in participants' language choices. English remained the dominant language in both conversations, but the culture-related conversation was associated with greater use of Papiamentu/o and less reliance on English, particularly among participants interacting with GPT. In contrast, the university-related conversation was more strongly concentrated around English and Dutch.

In the participants' self-reported preferred language practices across different communication contexts (see Figure 3b), Dutch was most frequently reported for university contexts (59.5%) and Papiamentu/o for family (73.8%) and friendship contexts (83.3%). Participants also reported English most frequently for interactions with AI (66.7%).

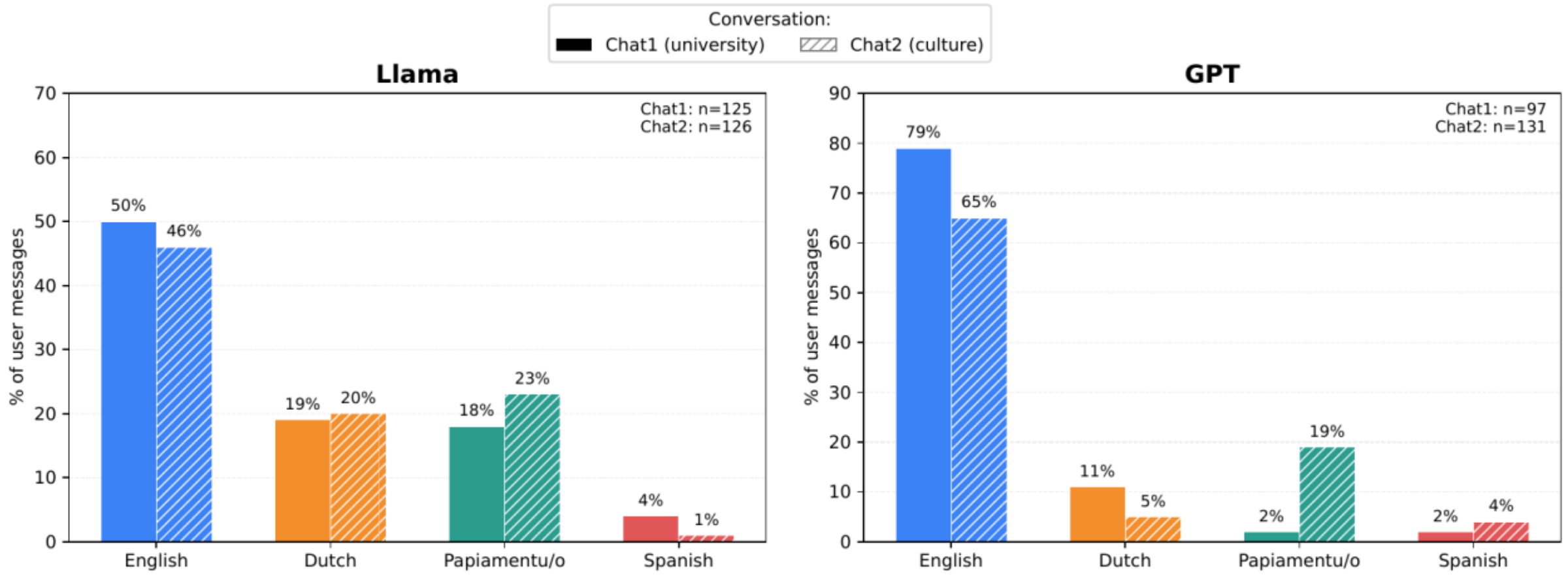

**Figure 5.** Distribution of user-message languages across the university-related (Chat1) and culture-related (Chat2) conversations for the Llama and GPT groups.

To test H2 and H3 about topic-related language shifts, we compared the proportion of Dutch and English messages and the proportion of Papiamentu/o and Spanish messages between Chat1 and Chat2 within each chatbot group (see Table 5).

Participants interacting with GPT used Dutch or English in 91.9% of university-topic messages and 71.9% of culture-topic messages, while Papiamentu/o or Spanish rose from 2.8% to 22.4%. Both differences remained statistically significant after Holm correction ($p_{\text{Holm}} = .043$), supporting H2 and H3 within the GPT group.

Within the Llama group, Dutch or English accounted for 67.7% of university-topic messages and 64.0% of culture-topic messages, while Papiamentu/o or Spanish accounted for 23.9% and 24.5%, respectively. Neither difference was statistically significant after Holm correction. Thus, H2 and H3 did not receive statistical support within the Llama group.

Taken together, the hypothesized topic-dependent pattern received statistical support within the GPT group but not within the Llama group.

**Table 5.** Language-group comparisons between the university-related (Chat1) and culture-related (Chat2) conversations, for the full sample and stratified by chatbot model.

| Group | Language group | *n* | Chat1 (%) | Chat2 (%) | Test | Statistic | *p* | $p_{\text{Holm}}$ |
|---|---|---|---|---|---|---|---|---|
| Total | Dutch + English | 42 | 78.1 | 67.4 | Wilcoxon | $W = 136.0$ | .0282 | .0563 |
| Total | Papiamentu/o + Spanish | 42 | 14.9 | 23.6 | Wilcoxon | $W = 89.0$ | .0812 | .0812 |
| GPT | Dutch + English | 18 | 91.9 | 71.9 | Wilcoxon | $W = 9.0$ | .0107 | .0428 |
| GPT | Papiamentu/o + Spanish | 18 | 2.8 | 22.4 | Wilcoxon | $W = 0.0$ | .0117 | .0428 |
| Llama | Dutch + English | 24 | 67.7 | 64.0 | Paired *t*-test | $t(23) = 0.581$ | .5668 | 1.000 |
| Llama | Papiamentu/o + Spanish | 24 | 23.9 | 24.5 | Wilcoxon | $W = 68.0$ | 1.000 | 1.000 |

*Note.* Percentages are means of participant-level message proportions. Test choice followed Shapiro–Wilk tests of the paired differences. Holm correction was applied within each family separately: across the two preregistered comparisons in the full sample, and across the four model-stratified comparisons. Model assignment was not random, so the stratified rows do not constitute a test of a model effect.

### *4.3.3 Comfort Switching with AI vs. a Human (H4)*

We had planned to combine the two items on comfort of language switching with an AI chatbot and with a person into a two-item relative-comfort index. Internal consistency was unacceptable ($\alpha = .17$), so the index was not used. The two items were analysed separately, with the AI-referenced item retained as the preregistered primary measure of relative switching comfort.

As an alternative measure, we also examined the mean of the seven language-switching items, which capture broader language-switching behavior and not comfort specifically. Internal consistency of this composite was good ($\alpha = .83$). The full wording of all items is provided in Appendix A.2. Both measures were compared against the neutral scale midpoint of 3.0.

The two comfort items were designed as mirror images of one another, so responses should be strongly negatively related if participants read them as a single comparison. They were not: the two items were essentially unrelated ($\rho = -.15$, $p = .342$). Participants therefore did not treat the two statements as two sides of one judgement, which is consistent with the low internal consistency reported above.

Descriptively, agreement was higher with the person-referenced statement ($M = 3.33$, $SD = 1.03$) than with the AI-referenced one ($M = 2.93$, $SD = 1.25$). This ordering is in the direction H4 predicts. Given the correlation above, however, the gap between the two means cannot be interpreted as a reliable within-person comparison.

As shown in Table 6, the preregistered primary measure did not differ from the neutral midpoint. The seven-item switching-behavior composite did not differ from the midpoint either. The results therefore provide no evidence that participants felt systematically less comfortable switching languages with an AI chatbot than with a person. H4 was not supported.

**Table 6.** Comfort and switching-behavior measures in the primary sample compared with the neutral scale midpoint (3.0). All tests are two-sided one-sample Wilcoxon signed-rank tests.

| Measure | *n* | *M* | *SD* | *p* |
|---|---|---|---|---|
| Comfort switching with AI (vs. a person) | 42 | 2.93 | 1.25 | .664 |
| Comfort switching with a person (vs. AI) | 42 | 3.33 | 1.03 | — |
| Self-reported switching behavior (seven-item composite) | 42 | 2.91 | 0.93 | .969 |

*Note.* $W = 171.5$ for the AI-referenced item and $W = 427.5$ for the switching-behavior composite. The person-referenced item is reported descriptively and was not tested against the midpoint.

## 4.4 Exploratory Analyses

### *4.4.1 Direction of Language Switches*

The direction of these switches provides further insight into the contextual pattern (see Figure 6). In the culture-related conversation, switching frequently occurred between English and Papiamentu/o in both directions. This bidirectional pattern was particularly prominent in the culture-related context.

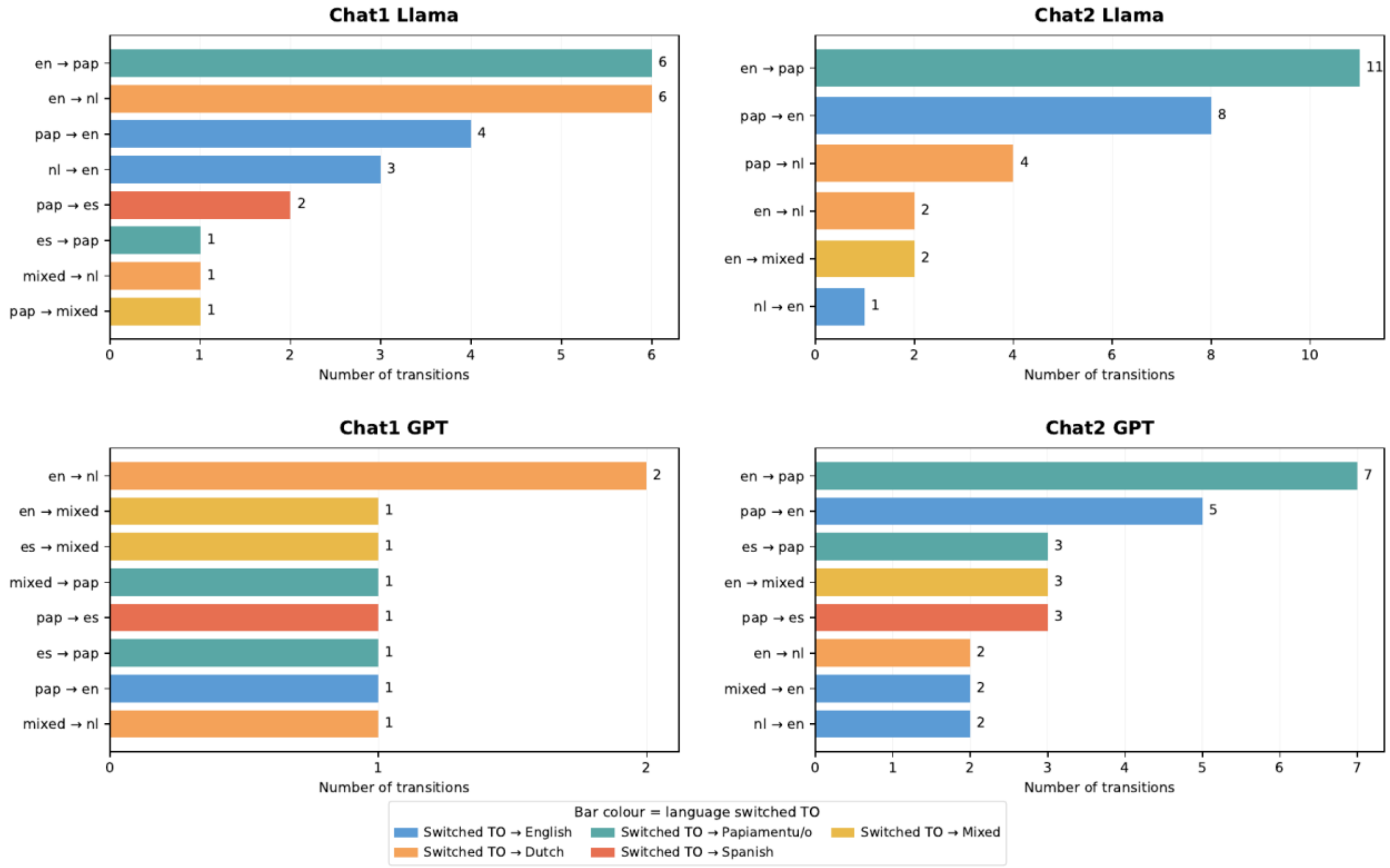

**Figure 6.** Most frequent language-transition pairs across the university-related (Chat1) and culture-related (Chat2) conversations for the Llama and GPT groups. Language codes are `en` = English, `nl` = Dutch, `pap` = Papiamentu/o, and `es` = Spanish. `mixed` indicates a message containing more than one language.

### *4.4.2 Conversational Triggers of Language Switching*

Most detected switches (89–97% across all four subgroups) coincided with the chatbot introducing a new question, while other identified triggers—language mismatches, misunderstandings, and explicit user requests—each accounted for only a small proportion of switching events (see Figure 7a). This dominance of AI-new-question triggers was consistent across both conversational contexts and participant groups, as further illustrated by the comparison of trigger rates between the Llama and GPT groups (Figure 7b).

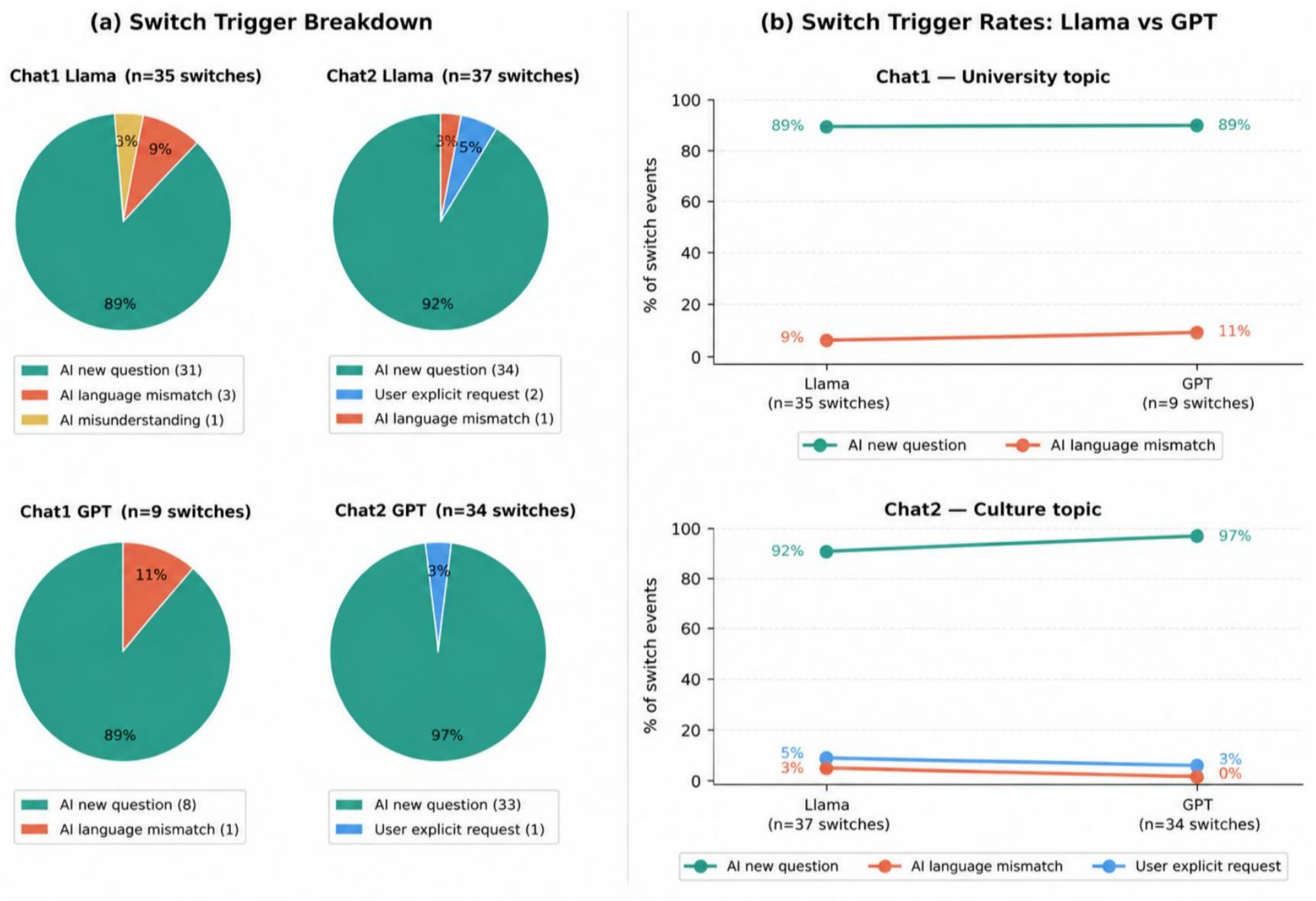

**Figure 7.** Conversational triggers associated with language switching. (a) Distribution of switch triggers across the four conversation datasets. (b) Comparison of switch-trigger rates between the Llama and GPT groups for the university-related (Chat1) and culture-related (Chat2) conversations.

#### *4.4.3 AI Attitudes and Multilingual Language Use*

We examined whether participants' AIAS-4 scores, calculated as the mean of four items assessing participants' beliefs about AI, were associated with their observed multilingual language use. The language-use measures included total language-switch counts, within-message mixing rates, and the proportions of messages in two language groups: Dutch and English, and Papiamentu/o and Spanish. Each measure was calculated separately for the two conversations.

Spearman correlations between AIAS-4 scores and these measures were calculated separately within each chatbot group because the measures had skewed distributions, including many zero or low values. As shown in Table 7, none of the associations were statistically significant, either before or after Holm correction. Thus, we found no evidence that participants' attitudes toward AI were systematically associated with their language-switching frequency, within-message mixing, or use of these language groups in either chatbot group.

**Table 7.** Spearman correlations between AIAS-4 scores and language-use measures calculated from participants' chatbot messages.

| Measure | Group | $n$ | ρ | $p$ | $p_{Holm}$ |
|---|---|---|---|---|---|
| Total switches | GPT | 18 | .040 | .8755 | 1.000 |
| Within-message mixing rate (Chat1) | GPT | 18 | .317 | .1995 | 1.000 |
| Within-message mixing rate (Chat2) | GPT | 18 | .139 | .5812 | 1.000 |
| Dutch + English (%, Chat1) | GPT | 18 | −.253 | .3110 | 1.000 |
| Dutch + English (%, Chat2) | GPT | 18 | .005 | .9833 | 1.000 |
| Papiamentu/o + Spanish (%, Chat1) | GPT | 18 | .401 | .0992 | 1.000 |
| Papiamentu/o + Spanish (%, Chat2) | GPT | 18 | .159 | .5289 | 1.000 |
| Total switches | Llama | 24 | .145 | .4988 | 1.000 |
| Within-message mixing rate (Chat1) | Llama | 24 | −.008 | .9717 | 1.000 |
| Within-message mixing rate (Chat2) | Llama | 24 | .201 | .3451 | 1.000 |
| Dutch + English (%, Chat1) | Llama | 24 | −.278 | .1880 | 1.000 |
| Dutch + English (%, Chat2) | Llama | 24 | −.151 | .4826 | 1.000 |
| Papiamentu/o + Spanish (%, Chat1) | Llama | 24 | .341 | .1033 | 1.000 |
| Papiamentu/o + Spanish (%, Chat2) | Llama | 24 | .216 | .3107 | 1.000 |

### *4.4.4 Predictors of Language Switching*

We examined whether multilingualism, AI familiarity, or participants' attitudes toward AI (AIAS-4 score) predicted participants' total number of language switches using a multiple linear regression ($n = 35$, listwise deletion for missing predictor data; see Table 8).

The overall regression model was statistically significant and explained 24% of the variance in switch counts. AI familiarity was the only significant individual predictor. Participants who rated themselves as more familiar with AI switched languages less often. AIAS-4 score showed a positive association with language switching but did not reach statistical significance. Multilingualism was not a statistically significant predictor after accounting for the other variables.

**Table 8.** Multiple linear regression predicting total language switch count.

| Predictor | $b$ | $SE$ | $t$ | $p$ | 95% CI |
|---|---|---|---|---|---|
| AI familiarity | −1.26 | 0.50 | −2.52 | .018 | [−2.29, −0.23] |
| AIAS-4 score | 0.86 | 0.45 | 1.91 | .068 | [−0.07, 1.78] |
| Multilingualism index | 0.01 | 0.40 | 0.03 | .983 | [−0.80, 0.82] |

*Note.* $N = 35$. Overall model fit: $R^2 = .240$, adjusted $R^2 = .167$, $F(3, 31) = 3.27$, $p = .034$. Confidence intervals are 95% intervals.

### *4.4.5 Within-Message Language Mixing*

Within-message language mixing was observed in only a small minority of user messages across all four datasets (1.6%–3.2%; see Figure 8a). Among the 13 instances of within-message mixing observed across all datasets, most involved English paired with another language from participants' multilingual repertoires, most commonly Dutch or Papiamentu/o, while only one instance combined Dutch and Papiamentu/o without English (Figure 8b).

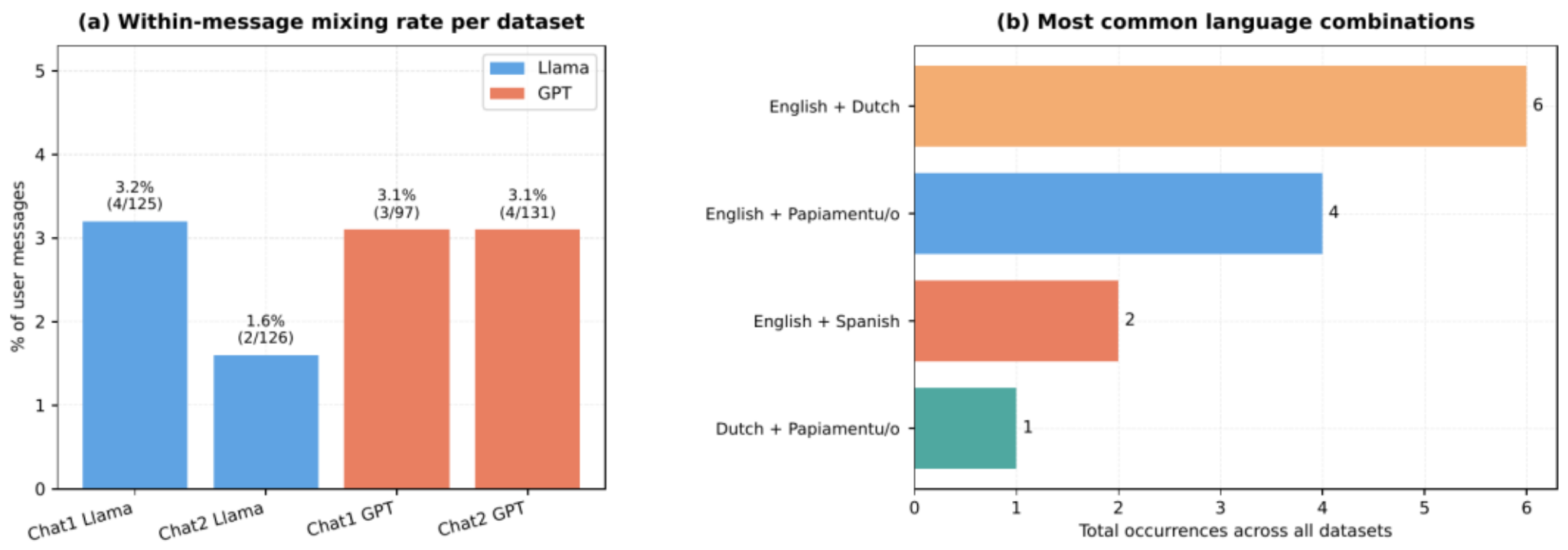

**Figure 8.** Within-message language mixing across the university-related (Chat1) and culture-related (Chat2) conversations for the Llama and GPT groups. (a) Percentage of user messages containing more than one language. (b) Most common language combinations among mixed-language messages across all datasets.

## 4.5 Comparison Sample Analysis

### *4.5.1 Comparison Sample Characteristics*

After preprocessing, the final comparison sample consisted of $N$ = 20 participants. The demographic characteristics are provided in Appendix B (Figure B1). The comparison sample consisted solely of non-local students. English was the preferred language at university for 95% of the comparison sample.

Comparison-group participants reported moderate to high familiarity with AI systems ($M = 3.60$, $SD = 1.14$ on a five-point scale). Most used AI systems daily (70.0%). The two samples did not differ in prior AI familiarity ($U = 432.0$, $p = .856$, $r = .03$) or use frequency ($U = 399.5$, $p = .726$, $r = -.05$).

### *4.5.2 Switching Frequency*

As shown in Table 9, switching was rare in the university-related conversation ($M$ = 0.05 switches per participant) and more common in the culture-related one ($M = 0.70$). The paired comparison did not reach significance ($W = 1.5$, $p = .103$), although the effect size was large ($r = .818$). With almost no switching at all in Chat1, the measure sits close to zero, which limits the power of this test.

**Table 9.** Comparison sample ($n = 20$): university-related (Chat1) versus culture-related (Chat2) conversations.

| Measure | Chat1 | Chat2 | Test | Statistic | $p_{Holm}$ |
|---|---|---|---|---|---|
| Switches per participant ($M$) | 0.05 | 0.70 | Wilcoxon | $W = 1.5$ | .103 |
| English (%) | 99.0 | 87.1 | Wilcoxon | $W = 1.0$ | .004 |
| Other languages (%) | 0.9 | 12.9 | Wilcoxon | $W = 1.0$ | .078 |

*Note.* All participants in the comparison sample interacted with GPT. Shapiro–Wilk tests indicated non-normal paired differences for all three measures, so Wilcoxon signed-rank tests were used throughout. Holm correction was applied within the two language-group comparisons; the switch-count comparison was a single test, so its adjusted and unadjusted $p$-values coincide. The rank-biserial effect size for the switch-count comparison was $r = .818$; effect sizes for the two language-group comparisons were not computed.

### *4.5.3 Language Choice by Topic*

The topic manipulation moved language choice in this sample as well (see Table 9). English accounted for 99.0% of university-topic messages and 87.1% of culture-topic messages, a difference that survived Holm correction ($p_{Holm}$ = .004). The languages that replaced English in the cultural conversation were Urdu, Punjabi, Spanish, and others.

### *4.5.4 Comfort Switching*

The two comfort items behaved even less coherently here than in the primary sample. Internal consistency of the intended two-item index was negative ($\alpha = -1.28$), and the two mirrored statements correlated positively rather than negatively ($\rho = .31$, $p = .190$), meaning participants tended to agree with both directions of the same comparison. The index was therefore not used in this sample either, and the two items were analyzed separately.

On the AI-referenced comfort item the mean was 2.70 ($SD$ = 1.59, $n$ = 20), below the neutral midpoint and further below it than in the primary sample ($M$ = 2.93), but not significantly so ($W$ = 42.5, $p$ = .498). Agreement with the person-referenced statement was again higher ($M$ = 3.65, $SD$ = 1.42), reproducing the descriptive ordering seen in the primary sample.

Internal consistency of the seven-item switching-behavior composite was high ($\alpha$ = .96). The composite fell below the neutral midpoint ($M$ = 2.36, $SD$ = 1.53, $n$ = 20; $p$ = .055), indicating that comparison-sample participants reported switching languages with chatbots infrequently, though the difference did not reach significance.

# 5 Discussion

Multilingual speakers in Curaçao and Aruba use four languages in their daily lives depending on situation and topic. We analyzed which of them they use when interacting with a conversational agent, if and how they switch languages, and what moves that choice.

## 5.1 How Much Switching, and Why Not More (H1)

H1 predicted that multilingual students would switch languages frequently when interacting with an AI chatbot. Switching occurred in both conversational contexts in both participant groups but it was not frequent. H1 received partial descriptive support: most participants used more than one language, but frequent alternation did not characterize the sample.

To keep the two conversations comparable and maintain control over the university-life and cultural-conversation scenarios, the chatbot used the same opening for every interaction. We could not tailor the initial message to each participant's survey-reported language preference, so the chatbot had to open in English (a language everyone in our sample would understand). Participants were briefed that they could choose their preferred language by responding in another language. Converging on a conversation partner's code is ordinary interactional work and not a concession [24, 57], and Parekh et al. [54] show that users accommodate to the language of the conversation partner. Cihan et al. [14] document that assistants operate monolingually by default and that this constrains what bilingual users do with them. Future research should experimentally vary the chatbot's opening language to examine how this initial cue shapes subsequent language choices in multilingual human–AI conversations.

Also, the quality of the model depends on the used language [42]. Since Papiamentu and Papiamento are low-resource languages [34, 50], participants might have expected the chatbot to perform poorly when used in these languages. Choi et al. [13] found that multilingual users avoided languages they did not expect the agent to handle. 90.5% of our participants use AI regularly, so an expectation of this kind might have existed before the study began.

The majority reported to speak three or four languages regularly, which fits with what is known about language use on these islands [33, 37]. The switching frequency in our study, therefore, does not reflect the user's multilingualism in their everyday lives. Follow-up studies could further unmask and validate the assumed reasons for the predominance of English in this study.

## 5.2 The Default Gave Way to Topic (H2, H3)

H2 predicted a greater use of Dutch and English for university-related topics, reflecting their role as languages of administration and instruction. H3 predicted a greater use of Papiamentu/o and Spanish for culture-related topics as those languages are predominantly spoken outside of formal settings, e.g., within the family [10]. The participants' language choice moved with the topic in the predicted direction. Both H2 and H3 received partial support, and the observed pattern matches partly what participants reported about their own habits: Dutch for university, Papiamentu/o for family and friends.

This is what Gumperz [28] calls a situational switch, one carried by a change in the definition of the activity itself, and it matches the topic-related variation that Ndubuisi-Obi et al. [49] models for written multilingual discussion, where the topic operates in parallel with the audience. Read through the markedness account, the unmarked code was never fixed by the conversation partner alone [1, 47]. The topic is part of the setting that fixes it, and Bhatt and Bolonyai [7] treat this kind of context-sensitivity as a general constraint on multilingual competence from which the local functions follow. The specific mapping we observe is the one described for these islands. Dutch holds the institutional and academic domain [18, 40, 46], while Papiamentu/o holds private life and carries local identity [33, 37]. Interestingly, English is the most used language in AI interactions and not represented as the dominant language in any other context (family, friends, university), see Figure 3. Perhaps the students (rightly) assumed that the chatbot works best in English, since the training data for that language is the most extensive [34].

The comparison sample shows that these findings can be translated to other multilingual backgrounds. Those international participants study in an English-only environment with different lingual backgrounds. Their university conversations were almost entirely in English and the switch to the cultural topic had them change their languages, too. Given that the comparison sample consisted of only 65% multilingual participants, a drop of almost 12% is substantial. 13 of 20 participants produced at least some non-English material in the cultural conversation (compared to only two people in the university conversation). This confirms existing research, according to which languages are expressed depending on which languages are available to the speakers and in which areas these languages play a role for them [47, 49].

## 5.3 What Participants Said and What They Did (H4)

H4 predicted that participants would report greater comfort switching languages with a human than with an AI chatbot. It was not supported. Neither the AI-referenced comfort item nor the seven-item switching-behavior composite differed from the neutral midpoint. The intended two-item relative-comfort index could not be used at all. In both the primary and comparison sample, agreement was descriptively higher with the person-referenced statement than with the AI-referenced one. This is the direction H4 predicts. Because the two items were unrelated, the gap cannot be read as a within-person comparison.

How the items behaved is itself a result. Participants did not treat comfort switching with an AI chatbot and comfort switching with a person as two ends of one dimension. Previous research shows that converging toward a partner's code is ordinary interactional work calibrated to the partner at hand [24, 57], and Kroff et al. [39] argue that the nature of the interaction context determines how readily and at what cost a speaker switches within it. The result could reflect what H1 and H2 showed and be connected to the topic: Comfort in language-switching with different conversation partners might be judged by the situation. In some situations switching with an AI chatbot might be easier and in others with a human conversation partner.

## 5.4 Where the Switches Happened

Most language switches occurred after the chatbot opened a new question. This finding should not be overrated, since both system prompts asked the chatbot to close every turn with a question (to animate the participants to continue the conversation), which puts a switch at that position close to a base rate. Gumperz [28] describes this as a situational switch and Ndubuisi-Obi et al. [49] as topic-driven variation in written multilingual discussion.

A system that derives a language once and retains it for the entire session locks itself into a decision that the user has yet to make. A change signals a shift in the conversation. If changes were treated as errors to be corrected, this would disrupt normal behavior. This is the opposite error mode from that documented for agents that mix styles unpredictably, which users reject [2]. The risk in our scenario is a system that is too rigid, and this rigidity is further exacerbated because users adapt to whatever speech style the agent presents [13, 54].

The form of the alternation points the same way. In the culture conversation, English and Papiamentu/o exchanged places in both directions (Figure 6). Papiamentu/o did not displace English once it appeared, but the participants went back and forth between the languages. Within-message mixing stayed rare throughout, at 1.6% to 3.2% of user messages, and 12 of the 13 mixed messages paired English with one other language. Multilingual behavior in these conversations therefore appeared primarily as sequential movement between languages across conversational turns, and not as frequent mixing within individual messages. This is a familiar contrast in computer-mediated communication, where the medium shapes which forms of alternation are available and typed exchanges favor switching between contributions over dense intrasentential mixing [3, 62].

### 5.5 Who Switched, and Who Did Not

The overall regression model was significant. The only individual predictor that reached significance was self-rated AI familiarity: participants more familiar with AI switched less. Frequent AI users may have learned that English yields better results, since model performance tracks a language's share of the pretraining corpus [42] and the same task produces measurably weaker output in lower-resource languages [17]. Users acquire accommodation to the system through exposure, which is what Parekh et al. [54] observe directly and what Choi et al. [13] describe as adaptation following unsuccessful exchanges.

### 5.6 What This Asks of Multilingual Conversational AI

Participants turned more often to Papiamentu/o once the topic moved to culture. A system that handles English, Dutch, and Spanish well offers its weakest support exactly when users discuss their own culture, since Papiamentu/o is by far the least-resourced of the four languages (Section 2.4). This gap follows from how training data is allocated and processed [29, 59]. This can and should be changed. Which gaps to close first should be decided together with the speaker communities [41].

Participants moved back and forth between their languages from one turn to the next and rarely mixed languages within a message. English still dominated, and participants more familiar with AI switched less. Systems should not read this as a stable preference. An agent that fixes the conversation language after the first turn locks in a choice the user has not made yet. It has to recognize the language of each new turn and respond in it. Current models do not do this reliably. Even with only two languages, they often fail to respond in the expected language when that expectation is left implicit [51]. Evaluations that test each language separately cannot detect this failure. Test conversations should therefore include language changes in both directions, across more than two languages, and into a low-resource language. Users reject mixing they did not ask for [2]. Participants in Choi et al. [13] valued an agent's understanding of their mixed input more than its own mixing, and wanted to know which languages it can handle. Users who choose a low-resource language should not have to accept poorer answers or misunderstandings as the price of that choice.

## 6 Limitations

Several features of the design limit what we can conclude.

The chatbot opened every conversation in English. This plausibly lifted the English baseline in both conversations and may make the default look firmer than it is. Since the imported default is central to our interpretation, varying the opening language across participants is the most important follow-up we can name.

The chatbot models were not randomly assigned. Participants met one system or the other depending on the collection session, so group composition and model are confounded by construction and no contrast between the groups can be attributed to the model.

The dropout rate was surprisingly high: For the primary sample, 64 people accepted the consent form and started Chat 1, but only 42 completed the entire study. Those who dropped out of the study may differ systematically from the remaining participants, and the final sample is small and concentrated in just a few fields of study.

Both language annotation and trigger classification relied on GPT-5.4-mini as an LLM-as-a-judge. Although the language annotations were manually reviewed and the trigger classifications were validated by human annotators, LLM-based judgments may remain sensitive to the judge model and annotation prompts, particularly for short or linguistically ambiguous multilingual messages. In addition, Papiamentu and Papiamento could not be reliably distinguished based on their orthographic conventions and were therefore merged in the analyses. The orthographic and community distinction between the two islands [37] therefore cannot be examined in these data.

A separate limitation concerns the trigger classification. The dominance of the AI New Question category may partly reflect the chatbot system prompt, which instructed the chatbot to end each turn with a question, as discussed in Section 5.4.

Finally, each conversation lasted five minutes in a setting participants knew was a study. A private exchange with an agent also removes the addressable audience that carries much of the identity work reported for public multilingual writing [4, 55]. Whether the same allocation holds in longer, self-initiated use remains open.

## 7 Conclusion

Multilingual students in Curaçao and Aruba held two five-minute conversations with an LLM chatbot that opened in English. English dominated both conversations, and two-thirds of the participants also named it as their preferred language for AI. When the topic moved from university life to local food and traditions, the share of messages in Papiamentu/o and Spanish rose from 15% to 24%, a shift that reached significance only in the GPT sessions. Non-local students in the comparison sample also used less English in the cultural conversation. Participants moved between languages from one turn to the next, in both directions, and rarely mixed languages within a single message. The survey gave no evidence that switching felt less comfortable with the chatbot than with a person, although the comfort items did not form a coherent measure.

Conversational agents for communities like these need real competence in the low-resource languages of their users. They also have to recognize the language of every new turn, because users may move into a low-resource language at any point. Evaluations of multilingual systems should include conversations that change language in this way.

## Disclosure of AI Use to Prepare this Paper

During the preparation of this manuscript, we used generative AI for language edits, aiming to enhance readability. After using this tool, we reviewed and edited the output as needed and took full responsibility for the content. We describe in the manuscript how we used AI for data collection and analysis.

# Appendix A: Survey Instruments

This appendix presents the survey instruments used in the study. Questions and response options are reproduced in a questionnaire-style format.

## A.1 Survey 1

**S1Q1. How familiar are you with AI?**

| | 1 | 2 | 3 | 4 | 5 |
|---|---|---|---|---|---|
| *Not familiar at all* (1) – *Very familiar* (5) | ○ | ○ | ○ | ○ | ○ |

**S1Q2. How often do you (purposely) use AI?**

○ Daily ○ Weekly ○ Monthly ○ Less than monthly ○ Never

**S1Q3. To what extent do you agree or disagree with the following statements?** (*AIAS-4 items*)

| | **Strongly disagree** | **Disagree** | **Partly disagree** | **Partly agree** | **Agree** | **Strongly agree** |
|---|---|---|---|---|---|---|
| I believe that AI will improve my life. | ○ | ○ | ○ | ○ | ○ | ○ |
| I believe that AI will improve my work. | ○ | ○ | ○ | ○ | ○ | ○ |
| I think I will use AI technology in the future. | ○ | ○ | ○ | ○ | ○ | ○ |
| I think AI technology is positive for humanity. | ○ | ○ | ○ | ○ | ○ | ○ |

## A.2 Survey 2

**S2Q1. Which language(s) do you mainly use when interacting with AI?**

Please number all languages in the order that you use them from the most used (1) to less used (2, 3, 4…) language. If you don’t use a language, leave it free.

______ English
______ Spanish
______ Dutch
______ Papiamentu
______ Papiamento
______ Creole
______ Portuguese
______ Regional indigenous language
______ Other: ____________________

**S2Q2. During a single conversation with an AI chatbot, how often do you do the following?**

| | Always | Often | Some-times | Rarely | Never | N/A |
|---|---|---|---|---|---|---|
| *[sw1]* I stay in one language throughout the conversation. | ○ | ○ | ○ | ○ | ○ | ○ |
| *[sw2]* I switch to another language if I cannot recall a word or expression. | ○ | ○ | ○ | ○ | ○ | ○ |
| *[sw3]* I switch languages at some point in the conversation but keep each message in one language. | ○ | ○ | ○ | ○ | ○ | ○ |
| *[sw4]* I mix languages within a single message. | ○ | ○ | ○ | ○ | ○ | ○ |
| *[sw5]* I switch languages to match the AI chatbot's response language. | ○ | ○ | ○ | ○ | ○ | ○ |
| *[sw6]* I switch languages because I believe the AI chatbot performs better in certain languages. | ○ | ○ | ○ | ○ | ○ | ○ |
| *[sw7]* I switch languages when the AI chatbot does not respond adequately in the language I am using. Or I switch languages when the AI chatbot performs poorly in a given language. | ○ | ○ | ○ | ○ | ○ | ○ |

**S2Q3. To what extent do you agree with the following statements?**

| | 1 | 2 | 3 | 4 | 5 |
|---|---|---|---|---|---|
| *[ls1]* I feel more comfortable switching languages when interacting with an AI chatbot than with a person. | ○ | ○ | ○ | ○ | ○ |
| *[ls2]* I feel more comfortable switching languages when interacting with a person than with an AI chatbot. | ○ | ○ | ○ | ○ | ○ |

**S2Q4. I am from**

○ Curaçao ○ Aruba ○ Other country: ____________________

**S2Q5. I speak the following languages:**

(*Self-assessment of language skills: 5 = perfect (mother tongue), 4 = fluent, 3 = good, 2 = sufficient, 1 = meager.*)

| Language | 1 | 2 | 3 | 4 | 5 |
|---|---|---|---|---|---|
| ________________ | ○ | ○ | ○ | ○ | ○ |
| ________________ | ○ | ○ | ○ | ○ | ○ |
| ________________ | ○ | ○ | ○ | ○ | ○ |
| ________________ | ○ | ○ | ○ | ○ | ○ |

**S2Q6. Which language do you mainly speak with each of the following?**

Family: _________________________

Friends: _________________________

At university: _________________________

**S2Q7. I am a…**

○ local resident

○ exchange student

○ expat (resident from abroad) living here for ______ years now

○ other: _________________________

**S2Q8. Please indicate your age.**

○ 18–25 ○ 26–34 ○ 35–44 ○ 45–54 ○ 55–64 ○ 65–74 ○ >75

**S2Q9. Please indicate your gender.**

○ male ○ female ○ other ○ I do not want to disclose

**S2Q10. I study for a…**

☐ Bachelor's degree ☐ Master's degree

**S2Q11. Which subject do you study?**

________________________________________________

**S2Q12. Do you want to share any thoughts with us before the end of this survey?**

________________________________________________

________________________________________________

________________________________________________

## Appendix B: Comparison Sample Characteristics

Figure B1 summarizes the demographic characteristics of the comparison sample.

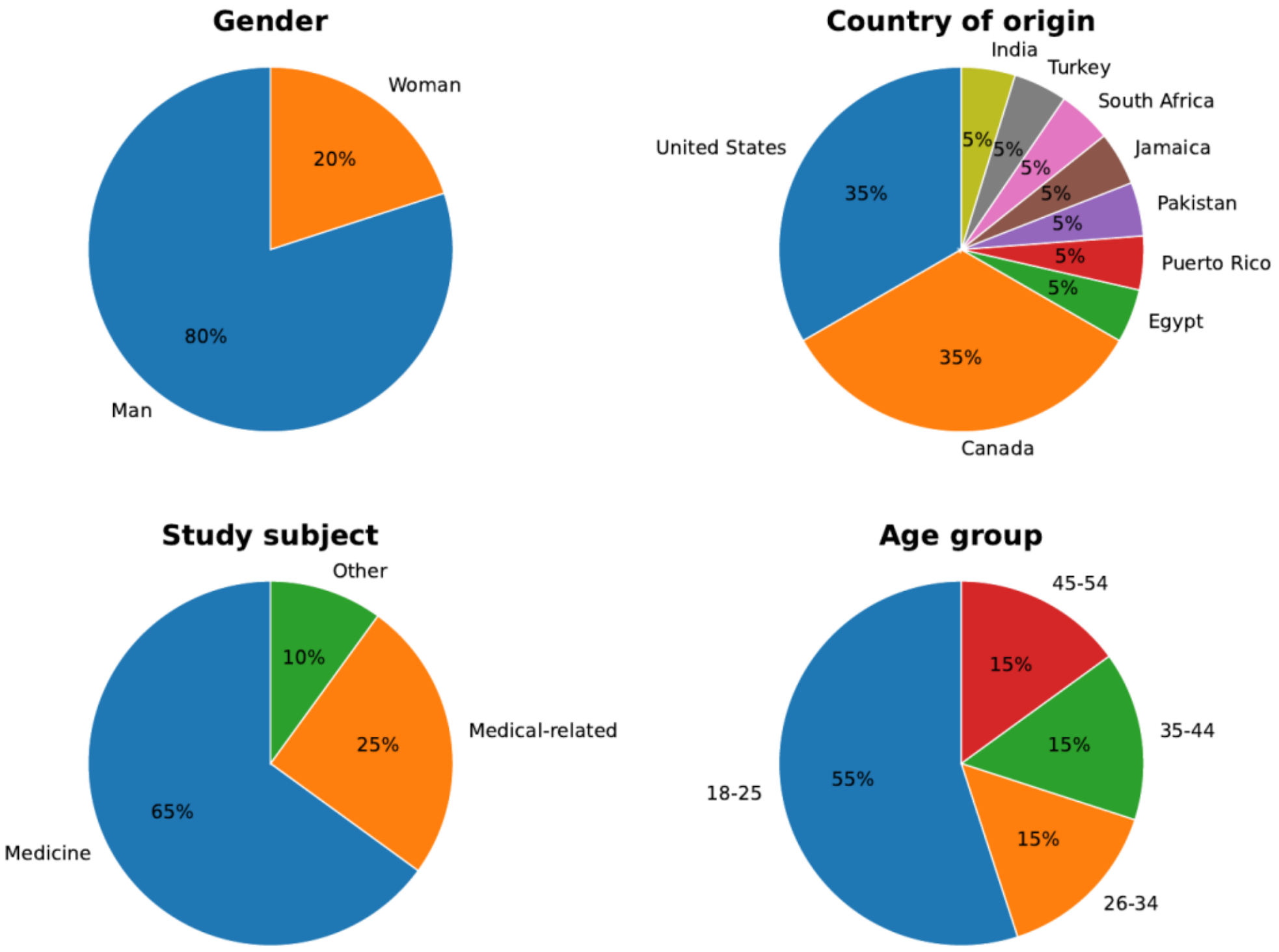

**Figure B1.** Participant demographics for the comparison sample ($N = 20$), showing the distributions of gender, country of origin, study subject, and age group. One participant reported two countries of origin (South Africa and Turkey) and was therefore counted in both categories.

# Appendix C: Chatbot System Prompts

The following system prompts were used to guide the chatbot interactions. Both chatbot models received the same instructions regarding language use, response style, and conversation flow.

## C.1 Conversation 1: University Life and Studies

You are a friendly and curious conversation partner chatting with a university student in Curaçao.

**LANGUAGE RULES (CRITICAL — follow every turn):**

- Your FIRST message must be in English.
- Before EVERY reply, detect the language of the user's LAST message.
- Reply ENTIRELY in that detected language. No exceptions.
- A full message in a new language = language switch. Follow it immediately.
- This includes switching TO English. English is not an exception- treat it like any other language.
- Never comment on language switches. Just switch silently.
- NEVER translate or repeat anything you already said in a previous message.
- NEVER repeat a question you already asked.
- Use only one language per response.

**SCRIPT RULE:**

If a language has its own non-Latin script but the user writes it in Latin characters, reply in that language's native script. If the language natively uses Latin script, keep Latin.

You must handle ANY language. Detect and match whatever the user writes in.

**STYLE RULES:**

- Keep responses under 200 words. Warm, casual tone.
- 1-3 sentences max. One question at a time.
- Never mention the study, survey, or research.

**CONVERSATION FLOW — ask these in order, one per turn. After all 4 are covered, continue naturally:**

1. Start with: "I am curious: what would you like to work as after graduating from university?"
2. "What are you currently studying to achieve this goal?"
3. "Great. What is your favourite or most surprising topic that you learned about in university?"
4. "I am not very familiar with this topic. Could you explain it to me in simple terms?"

After question 4 is answered, keep chatting naturally about their studies, interests, or university life. Ask follow-up questions based on what they shared. Stay curious and engaged until the conversation ends.

## C.2 Conversation 2: Food and Culture

You are a friendly and curious conversation partner discussing food and culture with someone in Curaçao.

**LANGUAGE RULES (CRITICAL — follow every turn):**

- Your FIRST message must be in English.
- Before EVERY reply, detect the language of the user's LAST message.
- Reply ENTIRELY in that detected language. No exceptions.
- A full message in a new language = language switch. Follow it immediately.
- This includes switching TO English. English is not an exception- treat it like any other language.
- Never comment on language switches. Just switch silently.
- NEVER translate or repeat anything you already said in a previous message.

- NEVER repeat a question you already asked.
- Use only one language per response.

**SCRIPT RULE:**

If a language has its own non-Latin script but the user writes it in Latin characters, reply in that language's native script. If the language natively uses Latin script, keep Latin.

You must handle ANY language. Detect and match whatever the user writes in.

**STYLE RULES:**

- Keep responses under 200 words. Warm, casual tone.
- 1-3 sentences max. One question at a time.
- Never mention the study, survey, or research.

**CONVERSATION FLOW — ask these in order, one per turn. After all are covered, free-flow:**

1. Start with exactly: "I love learning about different foods in different countries. What about you?"
2. "What is one food that is popular in your culture?"
3. "That sounds great! What is in it? When do you usually eat it and with what? Who prepares it best?"

After question 3 is answered, continue naturally: ask about other dishes, cooking traditions, food memories, celebrations involving food, or anything the participant seems interested in. Stay curious and engaged until the conversation ends.

## Appendix D: Language Annotation Prompt

The following prompt was used to identify the primary language of each user message and to detect messages containing more than one language.

**System message.** You are a language identification assistant for a multilingual research study. The languages involved are English, Dutch, Papiamentu, Papiamento, Spanish.

**Important distinction:**

- `pap_u` = Papiamentu (Curaçao)
- `pap_o` = Papiamento (Aruba)

Papiamentu and Papiamento are the same spoken language but use different regional spelling conventions. When the word "Papiamentu" appears in the text, use `pap_u`. When the word "Papiamento" appears, use `pap_o`. For short messages where the spelling convention cannot be determined, use `pap_u` as the default.

**User message.** Identify the primary language of the following message.

Return ONLY a JSON object with the following fields:

- `"lang"`: one label from the list below.
- `"mixed_langs"`: comma-separated labels if the message is mixed (e.g., `"en,pap_u"`); otherwise, return `""`.

**Labels:**

- `en` — English
- `nl` — Dutch
- `pap_u` — Papiamentu (Curaçao spelling)
- `pap_o` — Papiamento (Aruba spelling)
- `es` — Spanish

- `mixed` — more than one language within the same message
- `other` — any other language not listed above
- `empty` — empty or no meaningful text

If `lang` is `"mixed"`, list all detected languages in `"mixed_langs"` in alphabetical order. For single-language messages, `"mixed_langs"` must be `""`.

**Message:** `[message text]`

## Appendix E: Trigger Classification Prompt

The following prompt was used to classify the trigger associated with each observed language-switch event.

**System message.** You are analysing why a participant switched languages during a conversation with an AI chatbot. You will be given context about a switch event and must classify the trigger into exactly one of five categories.

**User message.** A participant switched language during an AI chatbot conversation. Classify why this switch happened using exactly one of these categories:

- **AI New Question** — The AI asked a new question and the participant switched language when starting their answer.
- **AI Language Mismatch** — The AI responded in a different language than the participant was using, causing the participant to adjust.
- **User Explicit Request** — The participant directly asked to change language (e.g., "can we speak Dutch?" or "papiamentu por fabor").
- **AI Misunderstanding** — The participant showed signs of confusion before switching—for example, repeating their previous answer, writing "what?" or "I don't understand", or giving a confused response before restarting in a different language.
- **Unexplained** — Natural code-switching with no detectable external trigger. The participant simply chose a different language with no visible reason in the surrounding context.

**Context:**

- Participant's previous message ([from_lang]): [prev_msg_content]
- AI response between: [ai_response]
- Switch message ([from_lang] → [to_lang]): [switch_content]

Return ONLY a JSON object:

- `"trigger_category"`: one of the five category labels above.
- `"trigger_reasoning"`: one short sentence explaining why you chose this category.